%% file: skeleton.tex
\documentclass{PoS}

\usepackage{subfig}
\usepackage[T1]{fontenc} 
\usepackage{graphicx}
\usepackage{epsfig}
\usepackage{axodraw2}
\usepackage{amsmath}
\usepackage{amssymb}
\usepackage{float}
\usepackage{placeins}
\usepackage{xcolor}
\usepackage{shuffle}
\usepackage{mathtools}
\usepackage{mathrsfs}
\usepackage{float}

\allowdisplaybreaks[4]

\newcommand{\be}{\begin{equation}}
\newcommand{\ee}{\end{equation}}
\newcommand{\nn}{\nonumber}
\newcommand{\beq}{\begin{equation}}
\newcommand{\eeq}{\end{equation}}
\newcommand{\bea}{\begin{eqnarray}}
\newcommand{\eea}{\end{eqnarray}}
\newcommand{\bfig}{\begin{figure}}
\newcommand{\efig}{\end{figure}}
\newcommand{\bc}{\begin{center}}
\newcommand{\ec}{\end{center}}

\newcommand{\D}{\mathcal{D}}
\newcommand{\intd}{\int\,\D^{d}k_{1}\D^{d}k_{2}\,}

\title{Two-loop non-planar master integrals for top-pair production in the quark-annihilation channel}

\ShortTitle{Non-planar master integrals for top-pair production
}

\author{\speaker{Matteo Becchetti}\\
        Center for Cosmology, Particle Physics and Phenomenology (CP3), Universit\'e Catholique de Louvain, 1348 Louvain-La-Neuve, Belgium\\
        E-mail: \email{matteo.becchetti@uclouvain.be}}
\author{Roberto Bonciani\\
        Dipartimento di Fisica, Sapienza - Universit\`a di Roma, Piazzale Aldo Moro 5, 00185, Rome, Italy\\
        E-mail: \email{roberto.bonciani@roma1.infn.it}}        
\author{Valerio Casconi\\
        Dipartimento di Fisica, Sapienza - Universit\`a di Roma, Piazzale Aldo Moro 5, 00185, Rome, Italy\\
        E-mail: \email{valerio.casconi@roma1.infn.it}}
\author{Andrea Ferroglia\\
        Physics Department, New York City College of Technology, The City University of New York, 300 Jay Street, Brooklyn, NY 11201 USA\\
        E-mail: \email{aferroglia@citytech.cuny.edu}}        
\author{Simone Lavacca\\
        Dipartimento di Fisica, Sapienza - Universit\`a di Roma, Piazzale Aldo Moro 5, 00185, Rome, Italy\\
        E-mail: \email{simone.lavacca@roma1.infn.it}}        
\author{Andreas von Manteuffel\\
        Department of Physics and Astronomy, Michigan State University, East Lansing, MI 48824, USA\\
        E-mail: \email{vmante@msu.edu}}

\abstract{We present the analytic computation of the master integrals associated to certain two-loop non-planar topologies, which are needed to complete the evaluation of the last two color coefficients for the top-pair production in the quark-annihilation channel, which are not yet known analytically. The master integrals have been computed exploiting the differential equations method in canonical form. The solution is given as a series expansion in the dimensional regularization parameter through to weight four, the expansion coefficients are given in terms of multiple polylogarithms.}

\FullConference{14th International Symposium on Radiative Corrections (RADCOR2019)\\ 
9-13 September 2019\\
		Palais des Papes, Avignon, France}

\begin{document}

\input{intro}

\input{notations}

\input{diffeq}

\input{conclusions}

\input{acknowledgments}

\bibliographystyle{JHEP}
\bibliography{biblio}

\end{document}

%% file: intro.tex
\section{Introduction \label{introduction}}

Top-pair production is one of the most studied processes at LHC and it provides crucial phenomenological results. As a consequence, theoretical predictions for this process are thoroughly studied. Numerical calculations for the total cross section and for differential distributions are known up to NNLO in perturbative QCD \cite{Baernreuther:2012ws,Czakon:2012zr,Czakon:2012pz,Czakon:2013goa,Czakon:2014xsa,Czakon:2015owf,Czakon:2016dgf,Czakon:2016ckf,Catani:2019iny}.

There are two relevant partonic channels for two-loop corrections to top-pair hadroproduction. The dominant production channel at LHC energies is the gluon-fusion channel $gg \to t\bar{t}$, while the quark-annihilation channel  $q\bar{q} \to t\bar{t}$ is the subdominant one. Two-loop contributions are described by the interference of the two-loop 
amplitude with the corresponding tree-level amplitude, they can be decomposed into a sum of independent color coefficients. For the  quark-annihilation channel the two-loop contributions to top-pair production are described by ten color coefficients, which are known numerically \cite{Czakon:2008zk} , while their infrared poles are known analytically \cite{Ferroglia:2009ep,Ferroglia:2009ii}. Moreover, eight of the color coefficients are also known analytically \cite{Bonciani:2008az,Bonciani:2009nb} in terms of multiple polylogarithms (MPLs) \cite{Goncharov:2001iea,Remiddi:1999ew,Vollinga:2004sn}.

Despite the fact that theoretical predictions are already known numerically, an analytic calculation of the two-loop contributions to top-pair production is still relevant. Indeed, it can be used as an independent check of the numerical results, and it could provide a faster and cheaper (in terms of CPU time) tool to evaluate the two-loop corrections needed in order to obtain phenomenological predictions for this process.

In this context, we report on the analytic computation of certain master integrals that are needed to evaluate the two color coefficients in the quark-annihilation channel which are not yet known analytically. Indeed, while part of the relevant master integrals have already been computed in other works \cite{Bonciani:2003te,Bonciani:2003hc,Bonciani:2008wf,Bonciani:2008az,Bonciani:2009nb,Bonciani:2010mn,vonManteuffel:2013uoa,Bonciani:2013ywa,Mastrolia:2017pfy} (see also the {\tt Loopedia} database~\cite{Bogner:2017xhp}), the non-planar master integrals that are described by the two non-planar integral families shown in figure \ref{fig1}\, have been computed just recently \cite{DiVita:2018nnh,Lee:2019lno,Becchetti:2019tjy,DiVita:2019lpl}.
In the present work, we summarise the results obtained for the master integrals of topology A \cite{Becchetti:2019tjy}, which have not been considered analytically so far, and we carry out an independent calculation for the master integrals of Topology B, originally evaluated in \cite{DiVita:2018nnh,Lee:2019lno}.
These results  will allow one to complete the analytic calculation of the two-loop corrections to top-pair production in the quark-annihilation channel.

The master integrals computation have been performed in dimensional regularization exploiting the differential equations method \cite{Kotikov:1990kg,Remiddi:1997ny,Gehrmann:1999as}. Employing Integration-By-Parts identities (IBPs) \cite{Tkachov:1981wb,Chetyrkin:1981qh,Laporta:2001dd}, it is possible to show that the scalar integrals associated to the two integral families taken into account, can be written in terms of a smaller set of master integrals. Specifically, the two topologies considered in this work involve  52 and 44 master integrals, respectively.
The reduction to master integrals has been performed by means of the computer programs
\textbf{FIRE} \cite{Smirnov:2008iw,Smirnov:2013dia,Smirnov:2014hma} and
\textbf{Reduze 2} \cite{Studerus:2009ye,vonManteuffel:2012np}, that implement integration-by-parts identities and Lorentz-invariance identities.
Then, we obtained analytic expressions in terms of MPLs for the master integrals, up to weight four in the dimensional regularisation parameter expansion, by solving the system of differential equations in canonical form \cite{Henn:2013pwa}. Finally, all the expressions have been checked numerically against the computer programs \textbf{SecDec} \cite{Carter:2010hi,Borowka:2012yc,Borowka:2015mxa,Borowka:2017idc} and \textbf{FIESTA} \cite{Smirnov:2008py,Smirnov:2013eza,Smirnov:2015mct}, which implement Sector Decomposition. Moreover, we stress that we also compared numerically our solutions against the expressions obtained by S. Di Vita, T. Gehrmann, S. Laporta, P. Mastrolia, A. Primo and U. Schubert \cite{DiVita:2019lpl}, who published results for a different choice of basis for the master integrals simultaneously to our work.

This contribution to the proceedings of RADCOR2019 is based on the work \cite{Becchetti:2019tjy}.

%% file: notations.tex
\section{Notations and computational setting \label{notations}}

In this section we describe the notations used and we introduce the computational framework. The two-loop QCD corrections to the top-pair production in the quark-annihilation channel can be described by the interference of the tree-level and two-loop amplitude, which admits the following color coefficients decomposition:
\bea
\label{qqColor}
A_{2L}^{q\bar{q}} & = & (N^2-1)\left[N^2 A^{q\bar{q}} + B^{q\bar{q}} + \frac{C^{q\bar{q}}}{N^2} + NN_l D^{q\bar{q}}_l + \frac{N_l}{N}E^{q\bar{q}}_l + NN_h D^{q\bar{q}}_h  \right. \nn \\
&& \left. + \frac{N_h}{N}E_h^{q\bar{q}} + N^2_l F_l^{q\bar{q}} + N_l N_h F_{lh}^{q\bar{q}} + N_h^2F_h^{q\bar{q}}\right].
\eea
The color coefficients ($A^{q\bar{q}},D^{q\bar{q}}_l,E^{q\bar{q}}_l,D^{q\bar{q}}_h,E_h^{q\bar{q}},F_l^{q\bar{q}},F_h^{q\bar{q}},F_{lh}^{q\bar{q}}$) are know analytically \cite{Bonciani:2008az,Bonciani:2009nb}, and it is possible to express them in terms of a well-studied class of functions, the \emph{Multiple polylogarithms} (MPLs) \cite{Goncharov:2001iea,Remiddi:1999ew}:
\begin{equation}
 G(a_{1},\dots,a_{n};z)=
 \int_{0}^{z}\frac{dt}{t-a_{1}} G(a_{2},\dots,a_{n};t) \, ,
 \end{equation}  
with
\begin{equation}
G(a_{1};z)=\int_{0}^{z}\frac{dt}{t-a_{1}} \quad \mbox{for} \quad a_{1}\neq 0,
\qquad \mbox{and} \quad G(\vec{0}_{n};z)=\frac{\log^{n}(z)}{n!} \, ,
\end{equation}
where $\vec{0}_{n}$ indicates a list of $n$ weights, all equal to $0$.
 On the other hand, an analytic expression for the coefficients $B^{q\bar{q}}$ and $C^{q\bar{q}}$ is not yet available, since they depend on certain Feynman integrals families whose analytic solution have been obtained just recently \cite{Mastrolia:2017pfy,DiVita:2018nnh,Becchetti:2019tjy,DiVita:2019lpl}.\par 
In the following we consider two non-planar integral families that contribute to $B^{q\bar{q}}$ and $C^{q\bar{q}}$ which have been computed in \cite{Becchetti:2019tjy}. They are described by the two  seven-denominator two-loop topologies, which we denote with the capital letters $A$ and $B$ (see Figure \ref{fig1}):
\begin{equation}
\label{eq:FamA}
\intd\frac{D_{4}^{-a_{4}}\,D_{6}^{-a_{6}}}{D_{1}^{a_{1}}\,D_{2}^{a_{2}}\,D_{3}^{a_{3}}\,D_{5}^{a_{5}}\,D_{7}^{a_{7}}\,D_{8}^{a_{8}}\,D_{9}^{a_{9}}} \, ,
\end{equation}
\begin{equation}
\label{eq:FamB}
\intd\frac{D_{5}^{-b_{5}}\,D_{6}^{-b_{6}}}{D_{1}^{b_{1}}\,D_{2}^{b_{2}}\,D_{3}^{b_{3}}\,D_{4}^{b_{4}}\,D_{7}^{b_{7}}\,D_{8}^{b_{8}}\,D_{9}^{b_{9}}} \, .
\end{equation}
The exponents $a_i$ and $b_i$, with $i=1,...,9$, are integer numbers where $a_4$, $a_6$, $b_5$, $b_6$ $\leq 0$, while the $D_i$, $i=1,...,9$, are the denominators and numerators involved, which are defined as:
\begin{eqnarray}
D_{\,i}&=&\lbrace -k_{1}^{2},\,-k_{2}^{2},\,-\left(p_{1}+k_{1}\right)^{2},\,-\left(p_{1}+k_{1}+k_{2} \right)^{2},\,-\left( k_{1}+p_{1}+p_{2}\right)^{2},\,-\left( k_{2}+p_{1}+p_{2}\right)^{2},\nn\\
&&-\left( k_{1}+k_{2}+p_{1}+p_{2}\right)^{2},\,m_t^2-\left(k_{1}+k_{2}+p_{3}\right)^{2},\,m_t^2-\left(k_{2}+p_{3}\right)^{2}
\rbrace \, .
\label{eq:dens}
\end{eqnarray}
We choose the following normalisation for the integration measure in eq.~\eqref{eq:FamA},\eqref{eq:FamB}:
\be
{\mathcal D}^dk_i = \frac{d^d k_i}{i \pi^{\frac{d}{2}}} e^{\epsilon\gamma_E} \left( \frac{m_t^2}{\mu^2} \right) ^{\epsilon} \, ,
\label{eq:intmeas}
\ee
where $\epsilon = (4-d)/2$ is the dimensional regularisation parameter, $\gamma_E$ is the Euler-Mascheroni constant and $\mu$ is the 't Hooft scale.  The kinematics of the process is described by the three Mandelstam invariants
\begin{equation}
s=(p_{1}+p_{2})^2,\quad t=(p_{1}-p_{3})^2, \quad u = (p_1-p_4)^2 \, ,
\end{equation}
which satisfy the relation $s+t+u=2m_t^2$.  The incoming quarks have momenta $p_1$ and $p_2$, while the final top quarks state have momenta $p_3$ and $p_4$.  All particles are on their mass-shell, namely $p_{1}^{2}\,=\,p_{2}^{2}=0$, and $p_{3}^{2}=p_{4}^{2}=m_t^2$, where $m_t$ is top-quark mass. The physical region is defined by 
\be
s > 4m_t^2 \, , \qquad t = m_t^2 - \frac{1}{2} \left( s-\sqrt{s(s-4m_t^2)} \cos{\theta} \right) \, ,
\ee
where $\theta$ is the scattering angle of top quark with respect to the direction of the incoming $q$ quark in the partonic center of mass frame.
\vspace{1cm}
%
\bfig
\bc
\[ 
\vcenter{
\hbox{
  \begin{picture}(0,0)(0,0)
\SetScale{1.0}
  \SetWidth{1}
\Line(-25,30)(-40,30)
\Line(-25,-30)(-40,-30)
\Line(-25,30)(-25,-30)
\Line(-25,30)(25,0)
\Line(0,-30)(25,30)
\Line(-25,-30)(25,-30)
%
%
  \SetWidth{4}
\Line(25,-30)(25,30)
\Line(25,30)(40,30)
\Line(25,-30)(40,-30)
\Text(0,-45)[c]{\footnotesize{(A)}}
\end{picture}}
}
\hspace*{5cm}
\vcenter{
\hbox{
  \begin{picture}(0,0)(0,0)
\SetScale{1.0}
  \SetWidth{1}
\Line(-25,30)(-40,30)
\Line(-25,-30)(-40,-30)
\Line(-25,30)(-25,-30)
\Line(-25,30)(25,0)
\Line(-25,0)(25,30)
\Line(-25,-30)(25,-30)
%
%
  \SetWidth{4}
\Line(25,-30)(25,30)
\Line(25,30)(40,30)
\Line(25,-30)(40,-30)
%
%
%
%
\Text(0,-45)[c]{\footnotesize{(B)}}
%
\end{picture}}
}
\]
\vspace*{7mm}
\caption{Seven-denominator topologies. Thin lines represent massless external particles and internal propagators, while thick lines represent massive external particles and internal propagators.
\label{fig1} }
\ec
\efig
%

Exploiting Integration-By-Parts Identities \cite{Tkachov:1981wb,Chetyrkin:1981qh,Laporta:2001dd}, it is possible to show that the scalar integrals associated to the two topologies $A$ and $B$ can be written in terms of 52 master integrals, for topology $A$, and 44 master integrals for topology $B$. However, since some master integrals are common to both topologies, the actual total number of master integrals is 70. For a subset of these master integrals analytic expressions in terms of MPLs were already been obtained previously
\cite{Bonciani:2003te,Bonciani:2003hc,Bonciani:2008wf,Bonciani:2008az,Bonciani:2009nb,Bonciani:2010mn,vonManteuffel:2013uoa,Bonciani:2013ywa,DiVita:2018nnh}, while for others, including the seven denominator four-point functions, analytic expressions are obtained for the first time \cite{Becchetti:2019tjy}.

\bfig
\bc
\includegraphics[scale=0.7]{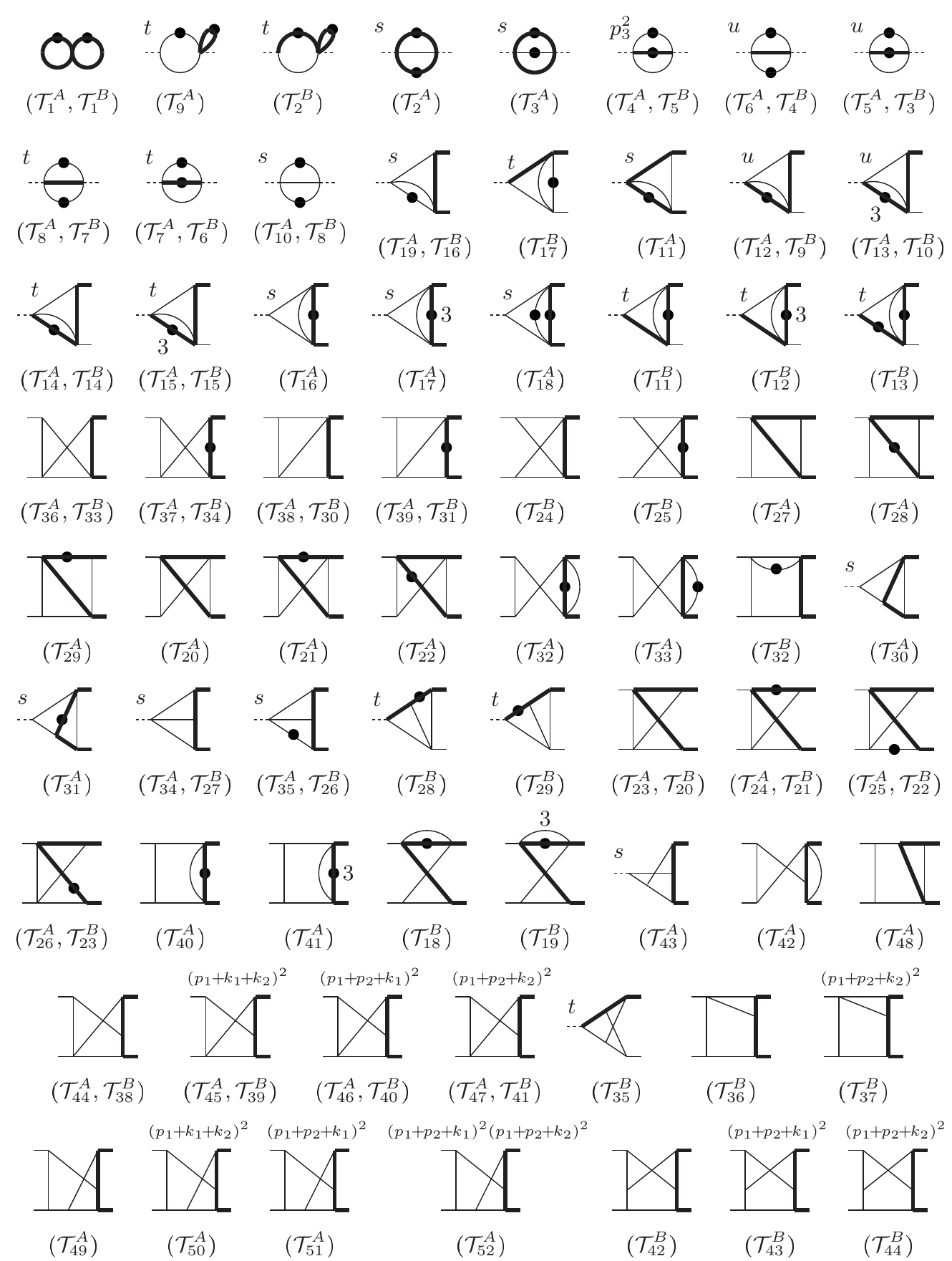}
\caption{Master Integrals in pre-canonical form. Internal thin lines represent massless propagators, while thick lines represent  heavy-quark (massive) propagators. External  thin lines represent massless particles on their mass-shell, $p^2=0$. External thick lines represent massive particles on their mass-shell, $p^2=m_t^2$. 
\label{fig2}}
\ec
\efig
%

%% file: diffeq.tex
\section{The Master Integrals}
\label{Sec3}

In this section we review the main steps for the computation of the master integrals associated to the integrals families $A$ and $B$.

\subsection{Differential Equations in Canonical Form}
 
In \cite{Becchetti:2019tjy} the master integrals computation has been performed exploiting the differential equation method  \cite{Kotikov:1990kg,Remiddi:1997ny,Gehrmann:1999as}, in particular it was possible to obtain a representation for the master integrals, up to weight four in $\epsilon$-expansion, in terms of MPLs by employing the Canonical Basis approach \cite{Henn:2013pwa},
 which consists in finding a basis for the MIs in which the system of differential equations has the specific form 
\be
\label{CanSys}
d\vec{f}(\vec{x},\epsilon) = \epsilon\, d\tilde{A}(\vec{x})\,\vec{f}(\vec{x},\epsilon) \, ,
\ee
where $\vec{f}(\vec{x},\epsilon)$ is the master integrals vector and $d$ is the total differential with respect to the kinematic invariants of the process $\vec{x}$.
The matrix $d\tilde{A}(\vec{x})$ in eq.~\eqref{CanSys} is a logarithmic differential one-form, i.e. $\tilde{A}$ has the form:
\be
\label{eq:dLog}
\tilde{A} = \sum_k A^{(k)} \log \alpha_k(\vec{x}),
\ee
where $A^{(k)}$ are matrices of rational numbers, and $\alpha_k(\vec{x})$ are algebraic functions of the kinematic invariants which determine the so called \emph{alphabet} of the process. In this basis the solution of the system of differential equations in \eqref{CanSys} is formally expressed in terms of Chen iterated integrals \cite{Chen:1977oja}:
\begin{equation}
\label{ChenInt}
\vec{f}(\vec{x},\epsilon)=\mathbb{P}\, \exp\left(\epsilon\int_{\gamma}d\tilde{A}(\vec{x}) \right)\vec{f}_{0}(\epsilon) \, ,
\end{equation} 
where $\mathbb{P}$ stands for the path-ordered integration, $\gamma$ is some path in the space of kinematic invariants and $\vec{f}_{0}(\epsilon)$ is a vector of boundary conditions.


\subsection{Roots rationalization and MPLs representation}

One of the main advantages of the canonical basis approach is that the formal solution \eqref{ChenInt} is particularly useful to obtain a representation for the master integrals as a series expansion in the dimensional regularization parameter $\epsilon$. Indeed, expanding around $ \epsilon \simeq 0$ eq.~\eqref{ChenInt} we get the following $\epsilon$-expansion for the master integrals vector:
\begin{equation}\label{eq:misexp}
    \vec{f}(\vec{x},\epsilon) = \vec{f}^{(0)}(\vec{x}_0) + \sum_{k=1}^\infty\epsilon^k \sum_{j=1}^k \int_0^1 dt_1 \frac{\partial\tilde{A}(t_1)}{\partial t_1} \int_0^{t_1} dt_2 \frac{\partial\tilde{A}(t_2)}{\partial t_2} \ldots \int_0^{t_{j-1}} dt_j \frac{\partial\tilde{A}(t_j)}{\partial t_j} \,\vec{f}^{(k-j)}(\vec{x}_0)\,,
\end{equation}
where $\vec{f}^{(i)}$ is the $i$-th term in the expansion and $t$ is some variable that specifies the integration path $\gamma = \gamma(t)$.\par
However, eq.~\eqref{eq:misexp} does not imply automatically that the master integrals $\vec{f}(\vec{x},\epsilon)$ admit a solution in terms of MPLs. Indeed, the explicit integration order-by-order in $\epsilon$ of eq.~\eqref{eq:misexp} in terms of MPLs is closely related to the class of functions to which the alphabet $\left\{\alpha_k(\vec{x})\right\}$ belongs to. Specifically, we have two different cases. If the alphabet $\left\{\alpha_k(\vec{x})\right\}$ is made by rational functions of the kinematic invariants, then it is possible to express the master integrals in terms of MPLs at any order in the $\epsilon$ expansion. Indeed, if the alphabet has a rational dependence on the kinematic invariants, then the matrix $d\tilde{A}$, which describes the system of differential equations in canonical form \eqref{CanSys},  has the form
\be
d\tilde{A} = \sum_i \sum_k A_{i,k}\frac{d\,x_i}{x_i - a_{i,k}}\, ,
\ee
where $a_{i,k}$ are algebraic functions of all kinematical invariants except for $x_i$. On the other hand, if the alphabet is not rational, i.e. it depends on roots of kinematic invariants, it is not trivial to obtain a representation in terms of MPLs by direct integration in eq.~\eqref{eq:misexp}. A possible solution to this problem is rationalize all the roots that appear in the differential equations system simultaneously by a suitable change of variables. This approach is not always available, indeed it can be proved that it is not always possible to rationalize simultaneously a set of roots, however we stress that, for the cases in which it is possible to rationalize all the roots, several methods have been proposed to apply this procedure \cite{Becchetti:2017abb,Besier:2018jen,Besier:2019kco}.\par
As last remark on the topic, we would like to observe that even if it is not possible to rationalize all the roots in the differential equations system, this does not imply that a representation for the master integrals in terms of MPLs is not available. Indeed, it has been recently shown in several examples \cite{Heller:2019gkq} that if the system of differential equations is in d-logarithmic canonical form, then it is possible to achieve an MPLs representation for the master integrals even in the presence of unrationalizable roots.

For the process under study the system of differential equations depends on the following two roots:
\be \label{eq:sqrt}
\sqrt{s(s - 4m_t^2)}, \;\;\;\;\; \sqrt{m_t^2 s (m_t^2 - t)(s + t - m_t^2)}\, .
\ee
We managed to rationalize this set of square roots exploiting the method described in \cite{Becchetti:2017abb}.  It is convenient to work with the dimensionless variables
\be
x = - \frac{s}{m_t^2}, \;\;\;\;\;\;\; y = - \frac{t}{m_t^2}\, ,
\ee
then in terms of $x$ and $y$ the change of variables that rationalizes \eqref{eq:sqrt} is given by:
\bea
x &\to& \frac{16w^2(1+4z)^2(w+z+4wz)^2}{z(1+8w)(z^2-4w^2)(z+4w+8wz)} \, , \nonumber \\ 
y &\to& 8 z(1+2z) \, . \label{eq:oldvariablechange}
\eea
However, this change of variables leads to long expressions for the solution of the master integrals, therefore we choose a different parametrization \cite{DiVita:2018nnh} to rationalize the square roots in eq.~\eqref{eq:sqrt}:
\be\label{eq:goodvar}
x = \frac{(1 - w)^2}{w}, \;\;\;\;\; y = \frac{1 - w + w^2 - z^2}{z^2 - w}.
\ee
Exploiting this change of variables, we are able to obtain more compact expressions for the master integrals in terms of MPLs, nevertheless we stress that the change of variables \eqref{eq:oldvariablechange} can be obtained in a fully algorithmic way by approaching the roots rationalization as a diophatine problem \cite{Becchetti:2017abb}.

In terms of the variables \eqref{eq:goodvar} the alphabet of the process is
\begin{align}
\begin{split}
\left\{ \alpha_k \right\} =  &\left\{ w, w - 1, w + 1, z , z - 1, z + 1, w - z,  w + z, w - z^2, w^2 - w + 1 - z^2, \right.  \\
& \left. w^2 - z^2(w^2 - w + 1), w^2 - 3w + z^2 + 1\right\}.
\end{split}
\end{align}
As a consequence, the analytic solution for the master integrals contains MPLs of argument w and weights
\begin{align}
\begin{split}
\left\{0 , \pm 1,  \pm z,  z^2, \frac{1 \pm \sqrt{4 z^2 - 3}}{2}, \frac{z(z \pm \sqrt{4z^2 - 3})}{2(z^2 -1)}, \frac{3 \pm \sqrt{5 - 4z^2}}{2} \right\},
\end{split}
\end{align}
and MPLs of argument z and weights
\be
\left\{0, \pm 1, \pm i\right\}.
\ee

\subsection{Numerical chekcs}
 
We conclude this section by comment on the numerical checks performed in order to validate our solutions for the master integrals. First, we describe the physical and non-physical phase-space regions with respect to the $w$ and $z$ variables \cite{Becchetti:2019tjy}. In the non-physical region $s < 0$, $t < 0$, the new variables are related to the adimensional Mandelstam invariants $x$ and $y$ by the relations:
\be\label{eq:unphys}
w = \frac{\sqrt{x + 4} - \sqrt{x}}{\sqrt{x+4} + \sqrt{x}}\, , \;\;\;\; z = \sqrt{\frac{1 - w + w^2 + wy}{1 + y}},
\ee
with
\be
0 < w < 1, \;\;\;\; \sqrt{w} < z < \sqrt{1 - w + w^2}.
\ee
On the other hand, in the physical region
\be
s > 4\, m_t^2\, ,\;\;\;\; m_t^2 - \frac{s}{2} - \frac{1}{2}\sqrt{s(s - 4m_t^2)} < t < m_t^2 - \frac{s}{2} + \frac{1}{2}\sqrt{s(s - 4m_t^2)},
\ee
the dependence of $w$ and $z$ on $x$ and $y$ has the form:
\be\label{eq:phys}
w = \frac{\sqrt{x' - 4} - \sqrt{x'}}{\sqrt{x' - 4} + \sqrt{x'}} + i 0^+,  \;\;\;\; z = \sqrt{\frac{1 - w + w^2 + wy}{1 + y}},
\ee
with the constraint 
\be
0 < -w < z < 1.
\ee
In eq.~\eqref{eq:phys}, the variable $x' = \frac{s}{m_t^2} >0$ is introduced for convenience, and the positive imaginary part comes from the analytic continuation of $x$, which is given by the Feynman prescription:
\be
x = - \frac{s + i0^+}{m_t^2} \equiv - x' - i0^+.
\ee

The analytic expressions for the master integrals in \cite{Becchetti:2019tjy} have been checked numerically for different points in both the physical and non-physical regions, by means of Sector Decomposition, exploiting the softwares \textbf{SecDec} \cite{Carter:2010hi,Borowka:2012yc,Borowka:2015mxa,Borowka:2017idc} and \textbf{FIESTA} \cite{Smirnov:2008py,Smirnov:2013eza,Smirnov:2015mct}, while the numerical evaluation of our analytic expressions has been performed using the software \textbf{GiNaC} \cite{Bauer:2000cp}. In order to obtain the desired numerical precision we wrote some of the canonical master integrals as linear combinations of quasi-finite integrals \cite{vonManteuffel:2014qoa,vonManteuffel:2017myy}, which are integrals with, at worst, a single pole in $\epsilon$ coming from the Euler Gamma function prefactor in their Feynman parameter representation. As shown in \cite{vonManteuffel:2014qoa,vonManteuffel:2017myy}, this class of integrals is evaluated more efficiently by both \textbf{SecDec} and \textbf{FIESTA}.  Moreover, we compared numerically our solutions for the master integrals with the ones obtained in \cite{DiVita:2019lpl}, we found complete agreement.

%% file: conclusions.tex
\section{Conclusions}

In this contribution to the proceedings of RADCOR2019 we reviewed the computation of the master integrals needed to complete the analytic evaluation of the color coefficients $B^{q\bar{q}}$ and $C^{q\bar{q}}$, which appear in the interference between the two-loop and the tree-level amplitudes for the top-pair production in the quark-annihilation channel.

The master integrals have been computed in dimensional regularization exploiting the differential equations method. In particular, we managed to put the system of differential equations in canonical form, and to rationalize the roots of kinematic invariants. As a consequence we were able to obtain, in an efficient an algorithmic way, expressions for the master integrals in terms of MPLs up to weight four in the dimensional regularization parameter expansion. 

Finally, we checked numerically our results against both the numerical codes \textbf{SecDec} and \textbf{FIESTA}, and the expressions obtained by the authors of \cite{DiVita:2019lpl} for a different basis choice of master integrals.

%% file: acknowledgments.tex
\section{Acknowledgments}

The work of A.F. is supported in part by the National Science Foundation under Grant No. PHY-1417354 and PSC CUNY Research Award TRADA-61151-00 49. The work of A.v.M. is supported in part by the National Science Foundation under Grant No. 1719863.